\documentclass[twocolumn,showpacs,showkeys,superscriptaddress]{revtex4}
\usepackage{graphicx}
\usepackage{bm}
\usepackage{color}
\usepackage{amsmath}
\usepackage{amsfonts}
\usepackage{amssymb}
\usepackage{natbib}
\usepackage{latexsym}
\usepackage{mathrsfs}

\begin{document}

\title{Magnetic Reconnection as a Mechanism for Energy Extraction \\ from Rotating Black Holes}
\author{Luca Comisso}
\email{luca.comisso@columbia.edu}
\affiliation{Department of Astronomy and Columbia Astrophysics Laboratory, Columbia University,
New York, New York 10027, USA.}
\author{Felipe A. Asenjo}
\email{felipe.asenjo@uai.cl}
\affiliation{Facultad de Ingenier\'{\i}a y Ciencias, Universidad Adolfo Ib\'a\~nez, Santiago 7941169, Chile.}


\begin{abstract}
Spinning black holes store rotational energy that can be extracted. When a black hole is immersed in an externally supplied magnetic field, reconnection of magnetic field lines within the ergosphere can generate negative energy (relative to infinity) particles that fall into the black hole event horizon while other particles escape stealing energy from the black hole. We show analytically that energy extraction via magnetic reconnection is possible when the black hole spin is high (dimensionless spin $a \sim 1$) and the plasma is strongly magnetized (plasma magnetization $\sigma_0 > 1/3$). The parameter space region where energy extraction is allowed depends on the plasma magnetization and the orientation of the reconnecting magnetic field lines. For $\sigma_0 \gg 1$, the asymptotic negative energy at infinity per enthalpy of the decelerated plasma that is swallowed by a maximally rotating black hole is found to be $\epsilon^\infty_-  \simeq - \sqrt{\sigma_0/3}$. The accelerated plasma that escapes to infinity and takes away black hole energy asymptotes the energy at infinity per enthalpy $\epsilon^\infty_+  \simeq \sqrt{3 \sigma_0}$. 
We show that the maximum power extracted from the black hole by the escaping plasma is $P_{\rm extr}^{\rm max} \sim 0.1 M^2 \sqrt{\sigma_0} \, w_0$ (here, $M$ is the black hole mass and $w_0$ is the plasma enthalpy density) for the collisionless plasma regime and one order of magnitude lower for the collisional regime. Energy extraction causes a significant spindown of the black hole when $a \sim 1$.
The maximum efficiency of the plasma energization process via magnetic reconnection in the ergosphere is found to be $\eta_{\rm max}  \simeq 3/2$. Since fast magnetic reconnection in the ergosphere should occur intermittently in the scenario proposed here, the associated emission within a few gravitational radii from the black hole is expected to display a bursty nature.
\end{abstract}


\maketitle

\section{Introduction}

Black holes are believed to play a key role in a number of highly energetic astrophysical phenomena, from active galactic nuclei to gamma-ray bursts to ultraluminous X-ray binaries. 
The extraordinary amounts of energy released during such events may have two different origins. It can be the gravitational potential energy of the matter falling toward an existing or forming black hole during accretion or a gravitational collapse. Or it can also be the energy of the black hole itself. Indeed, a remarkable prediction of general relativity is that a spinning black hole has free energy available to be tapped. How this occurs has fundamental implications for our understanding of high energy astrophysical phenomena powered by black holes.

It was shown by Christodoulou \cite{christodoulou70} that for a spinning (Kerr) black hole having mass $M$ and dimensionless spin parameter $a$, a portion of the black hole mass is ``irreducible'',
\begin{equation}
M_{\rm irr} =  M \sqrt{\frac{1}{2} \left( {1+\sqrt{1-a^2}} \right)}  \, .
\end{equation}
The irreducible mass has a one-to-one connection with the surface area of the event horizon, $A_H =4\pi(r_H^2+a^2) =  16 \pi M_{\rm irr}^2$, which is proportional to the black hole enthropy  $S_{\rm BH} = ({k_B c^3}/{4 G \hbar}) A_H$ \cite{bekenstein72,bekenstein73,hawking74,hawking75}, where $k_B$, $G$, $\hbar$, and $c$ denote, respectively, the Boltzmann constant, the gravitational constant, the reduced Planck constant, and the speed of light in vacuum. Thus, the maximum amount of energy that can be extracted from a black hole without violating the second law of thermodynamics is the rotational energy 
\begin{equation}
E_{\rm rot} = \left[ {1-\sqrt{\frac{1}{2} \left( {1+\sqrt{1-a^2}} \right)}} \right] M c^2 \, .
\end{equation}
For a maximally rotating black hole ($a =1$), this gives $E_{\rm rot} = (1-1/\sqrt{2}) M c^2 \simeq 0.29 M c^2$. Therefore, a substantial fraction of black hole energy can, in principle, be extracted \cite{note1}.

The possibility of extracting black hole rotational energy was first realized by Penrose \cite{penrose69}, who envisioned a thought experiment in which particle fission ($0 \rightarrow 1 + 2$) occurs in the ergosphere surrounding a rotating black hole. If the angular momentum of particle $1$ is opposite to that of the black hole and is sufficiently high, then the energy of particle $1$, as viewed from infinity, may be negative. Hence, since the total energy at infinity is conserved, the energy of particle $2$ as measured from infinity will be larger than that of the initial particle $0$. When the particle with  negative energy at infinity ($1$) falls into the black hole's event horizon, the total energy of the black hole decreases.  Therefore, the energy of the escaping particle $2$, which is higher than that of the original particle $0$, is increased at the expense of the rotational energy of the black hole.

Although the Penrose process indicates that it is possible to extract energy from a black hole, it is believed to be impractical in astrophysical scenarios. Indeed, energy extraction by means of the Penrose process requires that the two newborn particles separate with a relative velocity that is greater than half of the speed of light \cite{Bardeen_1972,wald74apj}, and the expected rate of such events is too rare to extract a sizable amount of black hole's rotational energy. On the other hand, Penrose's suggestion sparked the interest to find alternative mechanisms for extracting black hole rotational energy, such as superradiant scattering \cite{TP74}, the collisional Penrose process \cite{Piran75}, the Blandford-Znajek process \cite{BZ77} and the magnetohydrodynamic (MHD) Penrose process \cite{Takahashi90}.
Among them, the Blandford-Znajek process, in which energy is extracted electromagnetically through the magnetic field supported by an accretion disk around the black hole, is thought to be the leading mechanism for powering the relativistic jets of active galactic nuclei (AGNs) \citep[e.g.][]{McKGamm04,Hawley06,komissarov07,Tchekho11} and gamma-ray bursts (GRBs) \citep[e.g.][]{HKLee2000,Tchekho08,komissarov09}.

While different mechanisms of energy extraction have been carefully analyzed over the years, the possibility of extracting black hole rotational energy as a result of rapid reconnection of magnetic field lines has been generally overlooked. An exploratory study conducted by Koide and Arai \cite{KA} analyzed the feasibility conditions for energy extraction by means of the outflow jets produced in a laminar reconnection configuration with a purely toroidal magnetic field. In this simplified scenario, they suggested that relativistic reconnection was required for energy extraction, but the extracted power and the efficiency of the reconnection process were not evaluated. This is necessary for determining whether magnetic reconnection can play a significant role in the extraction of black hole energy. 
The recent advent of general-relativistic kinetic simulations of black hole magnetospheres \cite{parfrey} do indeed suggest that particles accelerated during magnetic reconnection may spread onto negative energy-at-infinity trajectories, and that the energy extraction via negative-energy particles could be comparable to the energy extracted through the Blandford-Znajek process.

In this paper we provide an analytical analysis of black hole energy extraction via fast magnetic reconnection as a function of the key parameters that regulate the process: black hole spin, reconnection location, orientation of the reconnecting magnetic field, and plasma magnetization.
Our main objective is to evaluate the viability, feasibility conditions, and efficiency of magnetic reconnection as a black hole energy extraction mechanism. 
In Section \ref{section2} we delineate how we envision the extraction of black hole rotational energy by means of fast magnetic reconnection, and we derive the conditions under which such energy extraction occurs. In Section \ref{section3} we show that magnetic reconnection is a viable mechanism of energy extraction for a substantial region of the parameter space.
In Section \ref{section4} we quantify the rate of energy extraction and the reconnection efficiency in order to evaluate whether magnetic reconnection is an effective energy extraction mechanism for astrophysical purposes. We further compare the power extracted by fast magnetic reconnection with the power that can be extracted through the Blandford-Znajek mechanism. Finally, we summarize our results in Section \ref{section5}.

\section{Energy Extraction by Magnetic Reconnection}  \label{section2}

\begin{figure}[]
\begin{center}
\includegraphics[width=8.5cm]{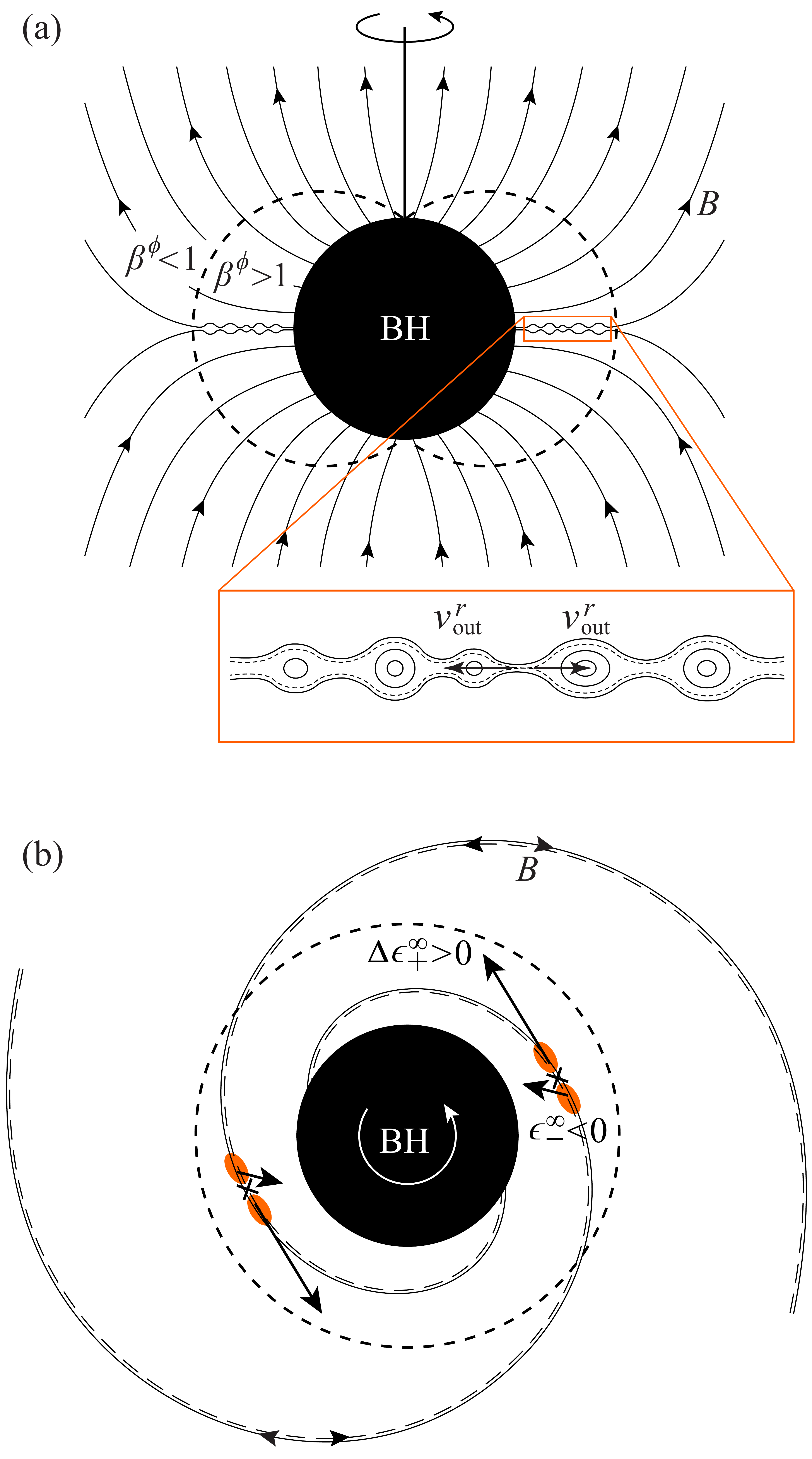}
\end{center}
\caption{Schematic illustration of the mechanism of energy extraction from a rotating black hole by magnetic reconnection in the black hole ergosphere. 
A configuration with antiparallel magnetic field lines adjacent to the equatorial plane is favored by the frame-dragging effect of the rapidly spinning black hole (panels (a) and (b) portray meridional and equatorial views, respectively), and the resulting equatorial current sheet is prone to fast plasmoid-mediated magnetic reconnection (see circular structures in the zoomed-in region \cite{noteplasmoids3D}). 
Magnetic reconnection in the plasma that rotates in the equatorial plane extracts black hole energy if the decelerated plasma that is swallowed by the black hole has negative energy as viewed from infinity, while the accelerated plasma with a component in the same direction of the black hole rotation escapes to infinity. 
The outer boundary (static limit) of the ergosphere is indicated by the short-dashed lines in both panels.  In panel (b), long-dashed and solid lines indicate magnetic field lines below and above of the equatorial plane, respectively. Finally, the dashed lines in the zoomed region indicate the two magnetic reconnection separatrices intersecting at the dominant magnetic reconnection $X$-point.}
\label{fig1}
\end{figure}

The possibility of extracting black hole rotational energy via negative-energy particles requires magnetic reconnection to take place in the ergosphere of the spinning black hole since the static limit is the boundary of the region containing negative-energy orbits. Magnetic reconnection inside the ergosphere is expected to occur routinely for fast rotating black holes. Indeed, a configuration with antiparallel magnetic field lines that is prone to magnetic reconnection is caused naturally by the frame-dragging effect of a rapidly spinning black hole. 
In this paper, we envision the situation illustrated in Fig. \ref{fig1}, where the fast rotation of the black hole leads to antiparallel magnetic field lines adjacent to the equatorial plane. 
This scenario is also consistent with numerical simulations of rapidly spinning black holes \citep[e.g.][]{parfrey,komissarov05,East18,ripperda20}.

The change in magnetic field direction at the equatorial plane produces an equatorial current sheet. 
This current sheet forms dynamically and is destroyed by the plasmoid instability (permitted by non-ideal magnetohydrodynamic effects such as thermal-inertial effects, pressure agyrotropy, or electric resistivity) when the current sheet exceeds a critical aspect ratio \cite{Comisso_2016,UzdLou_2016,Comisso_2017}. The formation of plasmoids/flux ropes (see circular sections in the zoomed-in region of Fig. \ref{fig1}) drives fast magnetic reconnection \citep[e.g.][]{daughton09,bhatta09}, which rapidly converts the available magnetic energy into plasma particle energy. 
Eventually, the plasma is expelled out of the reconnection layer and the magnetic tension that drives the plasma outflow relaxes. The field lines are then stretched again by the frame-dragging effect and a current layer prone to fast plasmoid-mediated reconnection forms again. This leads to reconnecting current sheets that form rapidly and intermittently.

Magnetic reconnection in the plasma that rotates around the black hole has the effect of accelerating part of the plasma and decelerating another part. If the decelerated plasma has negative energy at infinity and the accelerated one has energy at infinity larger than its rest mass and thermal energies (see the example regions in orange in Fig. \ref{fig1}(b)), then the plasma that escapes to infinity acquires energy at the expense of the black hole rotational energy when the negative-energy particles are swallowed by the black hole as in the standard Penrose process \cite{penrose69}. Therefore, we want to examine when magnetic reconnection in the ergosphere of the black hole redistributes the angular momentum of the plasma in such a way to satisfy these conditions. Furthermore, we want to evaluate if the extraction of black hole rotational energy via fast plasmoid-mediated reconnection can constitute a major energy release channel.

We describe the spacetime around the rotating black hole by using the Kerr metric in Boyer-Lindquist coordinates $x^\mu=(t, r, \theta, \phi)$, where $r$ is the radial distance, $\theta$ is the polar angle, and $\phi$ is the azimuthal angle. The Kerr metric can be expressed in terms of the square of the line element $d{s^2} = g_{\mu \nu} d{x^\mu}d{x^\nu}$ as \citep[e.g.][]{MTW}
\begin{equation} \label{BL_coord}
d{s^2} = g_{tt} d{t^2} + 2 g_{t\phi} dt d\phi  + g_{\phi\phi} d{\phi^2} + g_{rr} d{r^2}  + g_{\theta\theta} d{\theta^2}  \, ,
\end{equation}
where the non-zero components of the metric are given by
\begin{equation}
g_{tt} = \frac{2 Mr}{\Sigma} -1 \, , \; \; \; g_{t\phi} = - \frac{2 M^2 a r \sin^2 \theta}{\Sigma}  \, ,
\end{equation}
\begin{equation}
g_{\phi\phi} = \frac{A}{\Sigma} \sin^2 \theta  \, , \quad g_{rr} = \frac{\Sigma}{\Delta}  \, , \quad  g_{\theta\theta} = \Sigma  \, ,
\end{equation}
with
\begin{equation}
\Sigma  = {r^2} + {\left( {aM} \right)^2}{\cos ^2}\theta \, ,
\end{equation}
\begin{equation}
\Delta  = {r^2} - 2Mr + {\left( {aM} \right)^2} \, ,
\end{equation}
\begin{equation}
A = \big[ {{r^2} + {{\left( {a M} \right)}^2}} \big]^2 -  {\left( {aM} \right)^2} \Delta \, {\sin ^2}\theta \, .
\end{equation}
The only two parameters that appear in the metric are the black hole mass, $M$, and the black hole dimensionless spin, $0 \leq a \leq 1$. Here, and in all subsequent expressions, we use geometrized units with $G=c=1$.

The inner boundary of the ergosphere of the Kerr black hole, which coincides with the outer event horizon, is given by the radial distance 
\begin{equation}\label{outerevent}
r_{H}=M+ M ({1 - a^2})^{1/2} \, ,
\end{equation}
while the outer boundary (static limit) is given by 
\begin{equation}\label{outerergo}
r_{E}  = M+ M ({1- a^2 \cos^2 \theta})^{1/2} \,  ,
\end{equation}
which yields $r_{E} =2M $ at the equatorial plane $\theta=\pi/2$.
In this paper we make the simplifying assumption that magnetic reconnection happens in the bulk plasma that rotates circularly around the black hole at the equatorial plane. 
This corresponds to a Keplerian angular velocity 
\begin{equation}\label{keplerOmega}
\Omega_K= \pm \frac{M^{1/2}}{r^{3/2} \pm a M^{3/2}} \, ,
\end{equation}
as seen by an observer at infinity. The upper sign refers to corotating orbits, while the lower sign applies to counter-rotating orbits. Circular orbits can exist from $r \rightarrow \infty$ down to the limiting circular photon orbit, whose radius is given by
\begin{equation}\label{circularorbitphotonrad}
r_{\rm ph}=2M \left[ 1+\cos\left(\frac{2}{3} \arccos(\mp a) \right)\right] \, .
\end{equation}
For a maximally rotating black hole ($a =1$), one has $r_{\rm ph}=M$ (corotating orbit) or $r_{\rm ph}=4M$ (counter-rotating orbit). 
However, for $r > r_{\rm ph}$ not all circular orbits are stable. Non-spinning test particles can stably orbit the black hole if they are at distances larger than or equal to the innermost stable circular orbit \cite{Bardeen_1972} 
\begin{equation}\label{rmargbsc}
r_{\rm isco}=M\left[3+Z_2 \mp {\Big( {(3-Z_1)(3+Z_1+2Z_2)} \Big)^{1/2}}  \right] \, ,
\end{equation}
where 
\begin{equation}\label{}
Z_1=1+(1-a^2)^{1/3}[(1+a)^{1/3}+(1-a)^{1/3}] \, , 
\end{equation}
\begin{equation}\label{}
Z_2=(3a^2+Z_1^2)^{1/2} \, . 
\end{equation}
For a maximally rotating black hole $r_{\rm isco}=M$ (corotating orbit) or $r_{\rm isco}=9M$ (counter-rotating orbit). Here we focus on corotating orbits since we are interested in magnetic reconnection occurring inside the ergosphere. 
We also assume that the plasma acceleration through magnetic reconnection is localized in a small region (close to the dominant reconnection $X$-point) compared to the size of the black hole ergosphere.

In what follows, it is convenient to analyze the plasma energy density in a locally nonrotating frame, the so called ``zero-angular-momentum-observer'' (ZAMO) frame \cite{Bardeen_1972}. In the ZAMO frame, the square of the line element is given by $d{s^2} =  - d{{\hat t}^2} + \sum\nolimits_{i=1}^3 {{{(d{{\hat x}^i})}^2}}  = {\eta _{\mu \nu }}d{{\hat x}^\mu }d{{\hat x}^\nu }$, where
\begin{equation}
d\hat t = \alpha \, dt \, , \quad \; d{{\hat x}^i} = \sqrt{g_{ii}} \, d{x^i} - \alpha {\beta^i}dt \, 
\end{equation}
(no implicit summation is assumed over $i$), with $\alpha$ indicating the lapse function
\begin{equation}
\alpha= \left(  { -g_{tt} + \frac{g_{\phi t}^2}{g_{\phi\phi}} } \right)^{1/2}  =  \left(\frac{\Delta \Sigma}{A} \right)^{1/2}   \, 
\end{equation}
and $\beta^i$ indicating the shift vector $(0, 0, \beta^\phi)$, with
\begin{equation}
\beta^\phi =  \frac{\sqrt{g_{\phi\phi}} \, \omega^\phi}{\alpha} = \frac{\omega^\phi}{\alpha} \left(\frac{A}{\Sigma} \right)^{1/2} \sin\theta \, 
\end{equation}
and $\omega^\phi = - g_{\phi t}/g_{\phi\phi} = 2 M^2 a r/A$ being the angular velocity of the frame dragging. An advantage of this reference frame is that equations become intuitive since the spacetime is locally Minkowskian for observers in this frame. Hereinafter, quantities observed in the ZAMO frame are denoted with hats.
Vectors in the ZAMO frame are related to the vectors in the Boyer-Lindquist coordinates as $\hat b^{0}=\alpha b^{0}$ and $\hat b^{i}= \sqrt{g_{ii}} \, b^{i} - \alpha\beta^i b^{0}$ for the contravariant components, while $\hat b_{0}=b_{0}/\alpha + \sum\nolimits_{i=1}^3 {(\beta^i/\sqrt{g_{ii}}) \, b_i} $ and $\hat b_i= b_i/\sqrt{g_{ii}}$ for the covariant components.

We evaluate the capability of magnetic reconnection to extract black hole energy by examining the conditions for the formation of negative energy at infinity and escaping to infinity of the plasma accelerated/decelerated by the reconnection process in the ergosphere (in this work we do not address the origin of the plasma properties but rather assume a plasma with a given particle density and pressure). From the energy-momentum tensor in the one-fluid approximation, 
\begin{equation}
T^{\mu \nu} = p g^{\mu \nu} + w U^{\mu}  U^{\nu} + {F^\mu}_{\delta} F^{\nu \delta} - \frac{1}{4}  g^{\mu \nu} F^{\rho \delta} F_{\rho \delta} \, ,
\end{equation}
where, $p$, $w$, $U^{\mu}$, and $F^{\mu \nu}$ are the proper plasma pressure, enthalpy density, four-velocity, and electromagnetic field tensor, respectively, one has the ``energy-at-infinity'' density $e^\infty = - \alpha g_{\mu 0} T^{\mu 0}$. Therefore, the energy-at-infinity density is given by
\begin{equation}
e^\infty =  \alpha {\hat e} +  {\alpha \beta^\phi {\hat P}^\phi}    \, ,
\label{einfty}
\end{equation}
where 
\begin{equation}
{\hat e} = w \hat\gamma^2 -p  + \frac{{\hat B}^2 + {\hat E}^2}{2}  \, 
\end{equation}
is the total energy density and
\begin{equation}
{\hat P}^\phi = w \hat\gamma^2 {\hat v}^\phi + {\big({\bm{\hat{B}}} \times {\bm{\hat{E}}}\big)^\phi}   \, 
\end{equation}
is the azimuthal component of the momentum density, with  $\hat\gamma =  \hat U^0 = \big[ 1 -  \sum\nolimits_{i=1}^3 {{{(d{{\hat v}^i})}^2}}  \big]^{-1/2}$,  $\hat B^i = \epsilon^{ijk} \hat F_{jk}/2$, and $\hat E^i = \eta^{ij} \hat F_{j0} = \hat F_{i0}$.

The energy-at-infinity density can be conveniently separated into hydrodynamic and electromagnetic components as $e^\infty = e^\infty_{\rm hyd} + e^\infty_{\rm em}$, where 
\begin{equation}\label{enerhyd}
e^\infty_{\rm hyd}  = \alpha {\hat e}_{\rm hyd} + {\alpha \beta^\phi w \hat\gamma^2 {\hat v}^\phi }      \,  
\end{equation}
is the hydrodynamic energy-at-infinity density and
\begin{equation}\label{enerem}
e^\infty_{\rm em}  = \alpha {\hat e}_{\rm em} +  {\alpha \beta^\phi {\big({\bm{\hat{B}}} \times {\bm{\hat{E}}}\big)_\phi}  }   \,  
\end{equation}
is the electromagnetic energy-at-infinity density, with ${\hat e}_{\rm hyd} = w \hat\gamma^2 - p$ and ${\hat e}_{\rm em} = ({\hat B}^2 + {\hat E}^2)/{2} $ indicating the hydrodynamic and electromagnetic energy densities observed in the ZAMO frame. 
In this paper we assume an efficient magnetic reconnection process that converts most of the magnetic energy into kinetic energy, so that the electromagnetic energy at infinity is negligible with respect to the hydrodynamic energy at infinity. Then, from Eq. \eqref{enerhyd}, we can evaluate the energy-at-infinity density of the expelled plasma using the approximation that the plasma element is incompressible and adiabatic, which leads to \cite{KA}
\begin{equation}\label{enerhydincompress}
e^\infty_{\rm hyd}  = \alpha  \Big[  (\hat\gamma + \beta^\phi \hat\gamma {\hat v}^\phi)w -  \frac{p}{\hat\gamma}    \Big] \,  .
\end{equation}

To analyze the localized reconnection process, we introduce the local rest frame $x^{\mu \prime}=(x^{0 \prime}, x^{1 \prime}, x^{2 \prime}, x^{3 \prime})$ of the bulk plasma that rotates with Keplerian angular velocity $\Omega_K$ in the equatorial plane. We choose the frame $x^{\mu \prime}$ in such a way that the direction of $x^{1 \prime}$ is parallel to the radial direction $x^{1}=r$ and the direction of $x^{3 \prime}$ is parallel to the azimuthal direction $x^{3}=\phi$. The orientation of the reconnecting magnetic field lines is kept arbitrary as it ultimately depends on the large scale magnetic field configuration, the black hole spin, and is also time dependent. Indeed, the complex nonlinear dynamics around the spinning black hole induces magnetic field line stretching, with magnetic reconnection causing a topological change of the macroscopic magnetic field configuration on short time scales. 
Therefore, here we introduce the orientation angle 
\begin{equation}
\xi=\arctan \big({{v}_{\rm out}^{1 \prime}}/{{v}_{\rm out}^{3 \prime}} \big) \, ,
\label{anglexi}
\end{equation}
where ${{v}_{\rm out}^{1 \prime}}$ and ${{v}_{\rm out}^{3 \prime}}$ are the radial and azimuthal components of the  outward-directed plasma in the frame $x^{\mu \prime}$. Accordingly, the plasma escaping from the reconnection layer has velocities ${\bm{v}}_{\pm}^{\prime}=v_{\rm out} (\pm \cos\xi\, {\bm{e}}_3^{\prime} \mp \sin\xi\, {\bm{e}}_1^{\prime})$, with $v_{\rm out}$ indicating the magnitude of the outflow velocity observed in the frame $x^{\mu \prime}$ and the subscripts $+$ and $-$ indicating the corotating and counterrotating outflow direction, respectively. In the plasmoid-mediated reconnection regime, a large fraction of the plasma is evacuated through plasmoid-like structures \cite{noteplasmoids}, which can also contain a significant component of nonthermal particles. Such particles gain most of their energy from the motional electric field \citep[e.g.][]{GuoPoP20} and are carried out by the plasmoids (where most of them are trapped) in the outflow direction \citep[e.g.][]{sironi16}.

The outflow Lorentz factor $\hat\gamma$ and the outflow velocity component ${\hat v}^\phi$ observed by the ZAMO can be conveniently expressed in terms of the Keplerian velocity in the ZAMO frame and the outflow velocities in the local frame $x^{\mu \prime}$. From Eq. \eqref{keplerOmega}, we can express the corotating Keplerian velocity observed in the ZAMO frame as
\begin{equation}\label{keplerv}
\hat v_K = \frac{A}{\Delta^{1/2}} {\left[ {   \frac{ (M/r)^{1/2} -a (M/r)^2 }{r^3-a^2 M^3}    } \right]} -\beta^\phi \, .
\end{equation}
Then, using ${\hat v}_{\pm}^\phi = ({\hat v_K} \pm v_{\rm out} \cos \xi)/(1 \pm {\hat v_K} v_{\rm out} \cos \xi)$ for the azimuthal components of the two outflow velocities and introducing the Lorentz factors $\hat\gamma_K=(1-\hat v_K^2)^{-1/2}$ and $\gamma_{\rm out} =(1-v_{\rm out}^2)^{-1/2}$, we can write the energy-at-infinity density of the reconnection outflows as 
\begin{eqnarray}\label{energuis}
e^\infty_{{\rm hyd},\pm}& \!=\! &\alpha \hat\gamma_K \Bigg[ \left(1 \!+\! \hat v_K  \beta^\phi  \right) \gamma_{\rm out} w    \nonumber \\ 
&&  \pm \cos\xi  \left(\hat v_K \!+\!  \beta^\phi \right) \gamma_{\rm out} v_{\rm out} w  \nonumber \\ 
&& -\frac{p}{\left(1 \!\pm\! \cos\xi \, \hat v_K  v_{\rm out} \right) \gamma_{\rm out}  \hat\gamma_K^2}  \Bigg] \, , 
\end{eqnarray}
where the subscripts $+$ and $-$ indicate the energy-at-infinity density associated with corotating (${\bm{v}}_{+}^{\prime}$) and counterrotating (${\bm{v}}_{-}^{\prime}$) outflow directions as observed in the local frame $x^{\mu \prime}$.

The outflow velocity $v_{\rm out}$ can be evaluated by assuming that the local current sheet at the dominant $X$-point has  a small inverse aspect ratio $\delta_X /L_X \ll 1$, where $\delta_X$ and $L_X$ are the half-thickness and half-length of this local current sheet.
If we consider that the rest frame rotating with Keplerian velocity is in a gravity-free state and neglect general relativistic corrections  \cite{AsenjComisPRL,comiAsenjblackhole,AsenjComiPRD19}, then, the conservation of momentum along the reconnection neutral line gives
\begin{equation} 
w \gamma_{\rm out}^2 v_{\rm out}^2/L_X + {{B}_{\rm up}^2} \delta_X^2/L_X^3 \simeq ({{B}_{\rm up}}/\delta_X) ({{B}_{\rm up}} \delta_X/L_X) \, ,
\label{mom_eq}
\end{equation} 
where $B_{\rm up}$ is the local magnetic field strength immediately upstream of the local current sheet. Here we have used Maxwell's equations to estimate the current density at the neutral line in addition to the outflow magnetic field strength \cite{Lyubarsky,comiAsenjoPRLspecial}. We also assumed that the thermal pressure gradient force in the outflow direction is small compared to the magnetic tension force, as verified by numerical simulations of relativistic reconnection with antiparallel magnetic fields \cite{Liu17}. Then, from Eq. \eqref{mom_eq} one gets
\begin{eqnarray}\label{velocityBup}
v_{\rm out} \simeq \left[ {\frac{ \left( 1-\delta_X^2/L_X^2 \right) \sigma_{\rm up}}{1 + \left( 1-\delta_X^2/L_X^2 \right) \sigma_{\rm up}}} \right]^{1/2} \, ,
\end{eqnarray}
where $\sigma_{\rm up} = B_{\rm up}^2/w_0$ is the plasma magnetization immediately upstream of the local current sheet at the dominant $X$-point. Consequently, for $\delta_X /L_X \ll 1$, the outflow velocity reduces to $v_{\rm out} \simeq \left[ {{\sigma_{\rm up}}/{(1 + \sigma_{\rm up})}} \right]^{1/2}$.

The local magnetic field $B_{\rm up}$ can be connected to the asymptotic macro-scale magnetic field $B_0$ by considering force balance along the inflow direction. 
In the magnetically dominated regime, thermal pressure is negligible, and the inward-directed magnetic pressure gradient force must be balanced by the outward-directed magnetic tension (the inertia of the inflowing plasma is negligible if $\delta_X /L_X \ll 1$). Then, from geometrical considerations one gets \cite{Liu17}
\begin{equation} 
B_{\rm up} = \frac{1- (\tan \varphi)^2}{1+ (\tan \varphi)^2}  B_0  \, ,
\label{drop_B_eq}
\end{equation} 
where $\varphi$ is the opening angle of the magnetic reconnection separatrix. Estimating $\tan \varphi \simeq \delta_X/L_X$, we have simply  
\begin{eqnarray}\label{velocityB0}
v_{\rm out} \simeq {\left( {\frac{\sigma_0}{1 + \sigma_0}} \right)^{1/2}} \, ,\quad \gamma_{\rm out} \simeq {\left( {1+\sigma_0} \right)^{1/2}} \, ,
\end{eqnarray}
where we have defined $\sigma_0 = B_0^2/w_0$ as the plasma magnetization upstream of the reconnection layer. Accordingly, in the magnetically dominated regime $\sigma_0 \gg 1$, the reconnection outflow velocity approaches the speed of light. We finally note that in the presence of significant embedding of the local current sheet, the scaling of the outflow velocity could be weakened with respect to $B_0$, while Eq. \eqref{velocityBup} remains accurate \cite{Liu17,sironi16}.

We must point out that in the plasmoid-mediated reconnection regime considered here, the continuous formation of plasmoids/flux ropes prevents the formation of extremely elongated ``laminar'' reconnection layers, thereby permitting a high reconnection rate \citep[e.g.][]{daughton09,bhatta09}. Depending on the plasma collisionality regime, plasmoid-mediated reconnection yields an inflow velocity (as observed in the frame $x^{\mu \prime}$)
\begin{equation}  \label{recvelocity}
v_{\rm in} = 
\begin{cases}
\mathcal{O}(10^{-2}) & {\rm for} \quad \delta_X > \ell_k  \; [44\!-\!47] \\
\mathcal{O}(10^{-1}) & {\rm for} \quad \delta_X \lesssim \ell_k   \; [42, 43] \, ,
\end{cases}
\end{equation}
where $\ell_k$ is the relevant kinetic scale that determines the transition between the collisional and collisionless regimes. The collisional regime is characterized by $\delta_X > \ell_k$, while the collisionless regime occurs if $\delta_X \lesssim \ell_k$. For a pair (${e^-} {e^+}$) dominated plasma, we have \cite{comiAsenjoPRLspecial} $\ell_k = \sqrt{\gamma_{{\rm th},e}} \, \lambda_e$, where $\lambda_e$ is the nonrelativistic plasma skin depth and ${\gamma_{{\rm th},e}}$ is the electron/positron thermal Lorentz factor. 
If there is also a significant ion component, then \cite{daughton09} $\ell_k = \sqrt{\gamma_{{\rm th},i}} \,  \lambda_i$, where $\lambda_i$ is the nonrelativistic ion inertial length and ${\gamma_{{\rm th},i}}$ is the ion thermal Lorentz factor. 
We emphasize that the reconnection rate is independent of the microscopic plasma parameters when magnetic reconnection proceeds in the plasmoid-mediated regime. In particular, plasmoid-mediated reconnection in the collisionless regime has an inflow velocity $v_{\rm in}$ that is a significant fraction of the speed of light, which potentially allows for a high energy extraction rate from the black hole (see Sec. \ref{section4}).

The expression for the energy at infinity associated with the accelerated/decelerated plasma as a function of the critical parameters ($a$, $r/M$, $\sigma_0$, $\xi$) can be finally obtained by substituting the magnetization dependence of the outflow velocity into Eq. \eqref{energuis}. Then, the hydrodynamic energy at infinity per enthalpy $\epsilon^\infty_\pm = e^\infty_{{\rm hyd},\pm}/w$ becomes
\begin{eqnarray}\label{energuisMagnet}
\epsilon^\infty_\pm& \!=\! &\alpha \hat\gamma_K \Bigg[ \left(1 \!+\! \beta^\phi \hat v_K\right) {\left( {1 \!+\! \sigma_0} \right)^{1/2}}  \pm \cos{\xi}  \left(\beta^\phi \!+\! \hat v_K \right) \sigma_0^{1/2} \nonumber\\
&&\qquad\qquad -  \frac{1}{4}  \frac{{\left( {1 \!+\! \sigma_0} \right)^{1/2}}  \mp \cos{\xi} \, \hat v_K \sigma_0^{1/2}}{\hat\gamma_K^2 (1+\sigma_0 \!-\! \cos^2{\xi} \, \hat v_K^2 \sigma_0)}\,  \Bigg]\, ,
\end{eqnarray}
where we have assumed a relativistically hot plasma with polytropic index $\Gamma=4/3$. 
Similarly to the original Penrose process \cite{penrose69}, energy extraction from the black hole through magnetic reconnection occurs when 
\begin{equation}\label{conditionsenergy}
\epsilon^\infty_-<0\,  \quad    {\rm and}   \quad   \Delta \epsilon^\infty_+ >0 \, ,
\end{equation}
where 
\begin{equation}\label{conditionsenergy2}
\Delta \epsilon^\infty_+ = \epsilon^\infty_+ - \left( {1-\frac{\Gamma}{\Gamma-1}  \frac{p}{w} } \right) =  \epsilon^\infty_+ 
\end{equation}
for a relativistically hot plasma.
Therefore, black hole rotational energy is extracted if the decelerated plasma acquires negative energy as measured at infinity, while the plasma that is accelerated acquires energy at infinity larger than its rest mass and thermal energies. 

\begin{figure}[]
\begin{center}
\vspace{0.20cm}
\includegraphics[width=8.4cm]{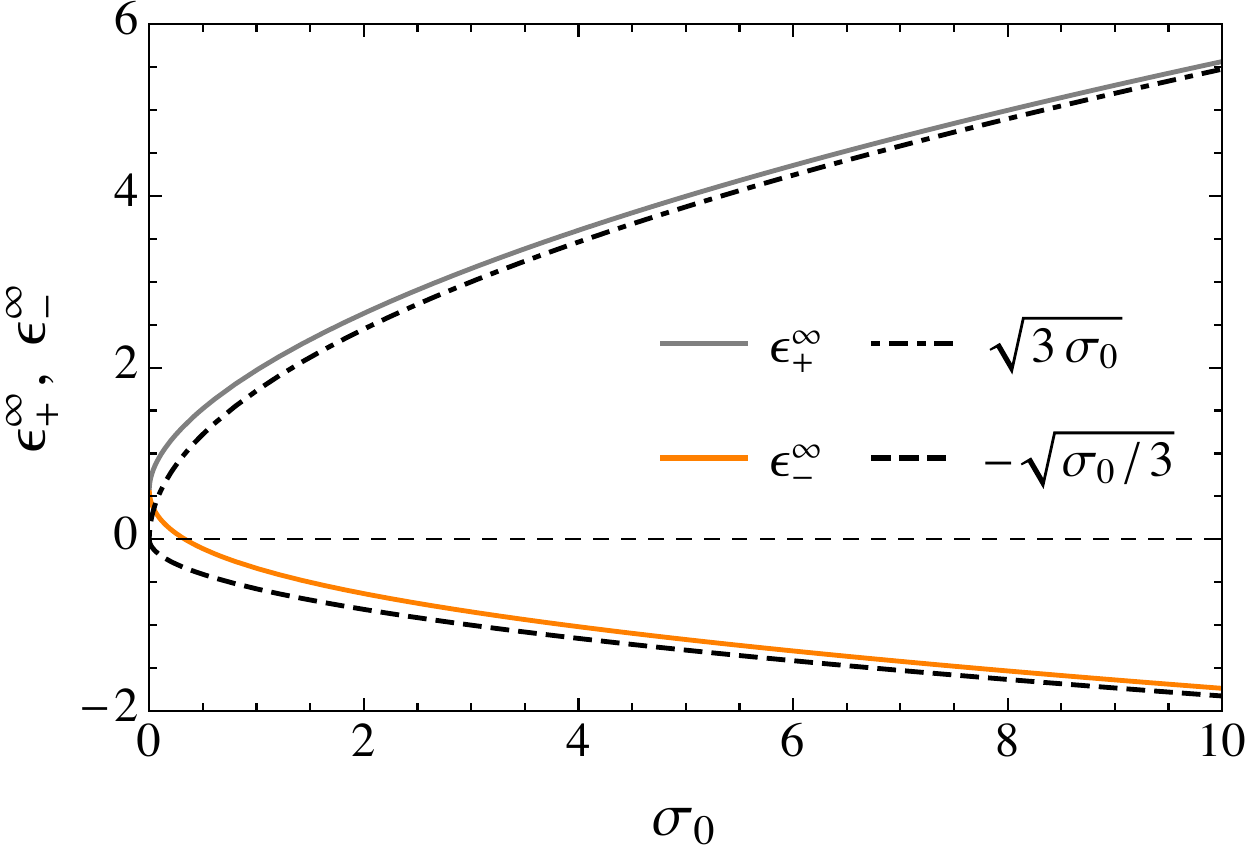}
\vspace{-0.30cm}
\end{center}
\caption{Energy at infinity per enthalpy $\epsilon^\infty_+$ (gray line) and $\epsilon^\infty_-$ (orange line) for optimal energy extraction conditions ($a, r/M \rightarrow 1$ and $\xi \rightarrow 0$). Energy extraction requires $\sigma_0 > 1/3$. For $\sigma_0 \gg 1$, $\epsilon^\infty_+ \simeq  \sqrt{3 \sigma_0}$ (dash-dotted black line) and $\epsilon^\infty_-  \simeq - \sqrt{\sigma_0/3}$ (dashed black line).}
\label{fig2}
\end{figure}

The energy at infinity per enthalpy $\epsilon^\infty_\pm$ given by Eq. \eqref{energuisMagnet} depends on the black hole spin $a$ and the $X$-point distance $r/M$, as well as the plasma magnetization $\sigma_0$ and the orientation angle $\xi$, which encodes the information of the magnetic field configuration surrounding the black hole. Eqs. \eqref{energuisMagnet}-\eqref{conditionsenergy2} indicate that energy extraction is favored by lower values of the orientation angle $\xi$ and higher values of the magnetization $\sigma_0$. It is instructive to consider the limit $a \rightarrow 1$, $\xi \rightarrow 0$, and $r \rightarrow M$ (the metric \eqref{BL_coord} has a coordinate singularity at the event horizon that can be removed by a coordinate transformation). In this case, from Eq. \eqref{energuisMagnet} we obtain $\epsilon^\infty_+>0$ and $\epsilon^\infty_-<0$  when 
\begin{equation}\label{}
\sigma_0 > {1}/{3}  \, .
\end{equation}
Therefore, in principle, it is possible to extract rotational energy via magnetic reconnection for values of $\sigma_0$ below unity. However, higher $\sigma_0$ values are required to extract sizable amounts of energy. If, in addition to $a, r/M  \rightarrow 1$ and $\xi \rightarrow 0$, we also consider $\sigma_0 \gg 1$, from Eq. \eqref{energuisMagnet} we obtain
\begin{equation}\label{energ_mas_simple}
\epsilon^\infty_+  \simeq  \sqrt{3 g_{\phi\phi}} \, {\omega^\phi} \gamma_{\rm out} v_{\rm out} \simeq  \sqrt{3 \sigma_0}  \, ,
\end{equation}
\begin{equation}\label{energ_minus_simple}
\epsilon^\infty_-  \simeq -  {\sqrt{\frac{g_{\phi\phi}}{3}} \, \omega^\phi} \gamma_{\rm out} v_{\rm out}  \simeq - \sqrt{\frac{\sigma_0}{3}}  \, .
\end{equation}
These relations give us the energy at infinity per enthalpy of the accelerated ($+$) and decelerated ($-$) plasma in the maximal energy extraction regime (as can be seen from Fig. \ref{fig2}, they provide a fairly accurate estimate already at values of $\sigma_0$ moderately larger then unity).

In the next sections, we will show that magnetic reconnection is a viable mechanism for extracting energy from rotating black holes for a significant region of the parameter space, we will evaluate the rate of black hole energy extraction, and we will determine the efficiency of the reconnection process.

\section{Energy Extraction Assessment in Phase Space} \label{section3}

\begin{figure}[]
\begin{center}
\includegraphics[width=8.4cm]{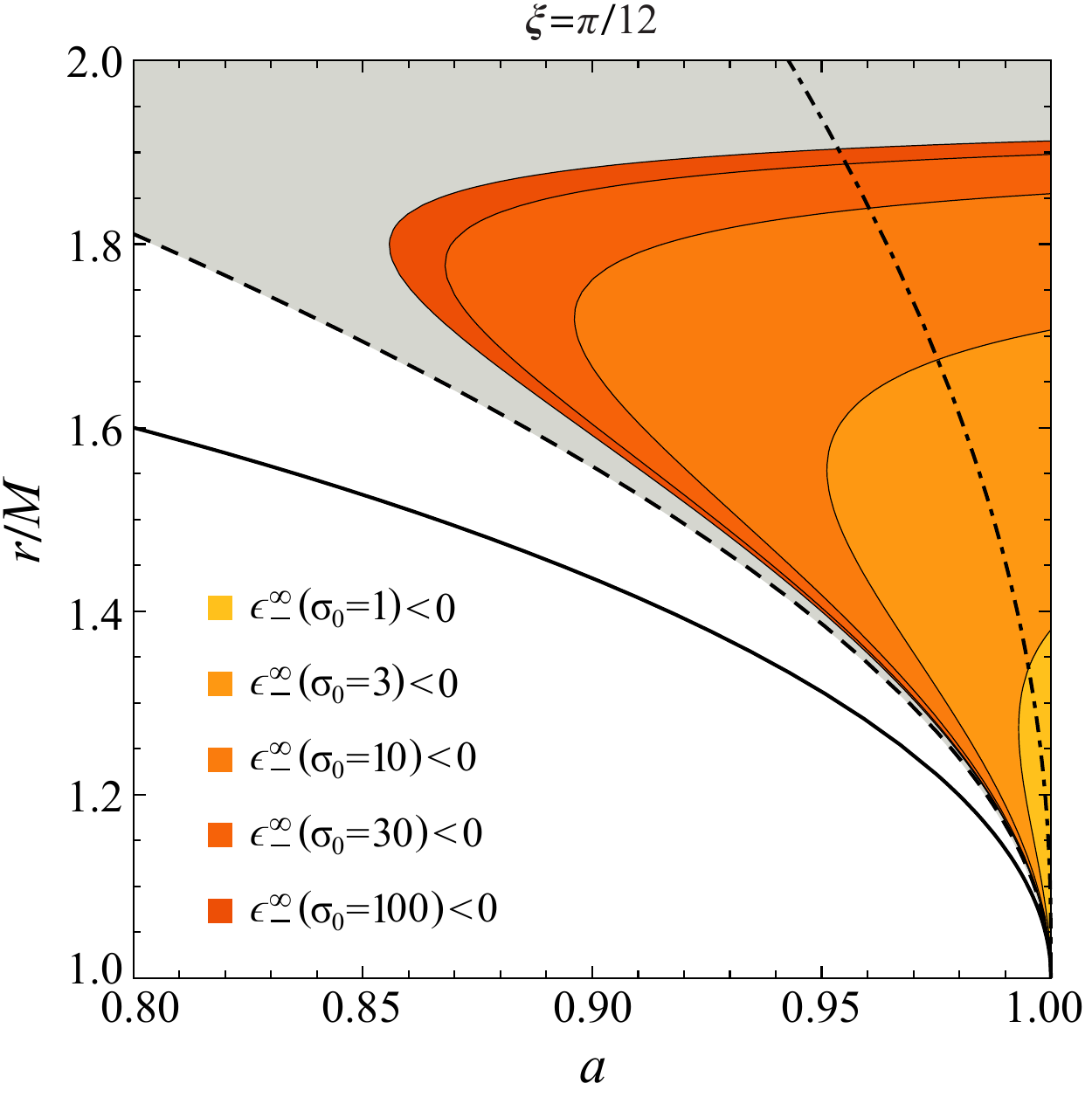}
\vspace{-0.30cm}
\end{center}
\caption{Regions of the phase-space $\{a,r/M\}$ where the energies at infinity per enthalpy from Eq.  \eqref{energuisMagnet} are such that $\Delta \epsilon^\infty_+ >0$ (gray area) and $\epsilon^\infty_- <0$ (orange to red areas), for a reconnecting magnetic field having orientation angle $\xi = \pi/12$ and different values of the magnetization parameter $\sigma_0 \in \left\{ {1,3,10,30,100} \right\}$. The area with $\epsilon^\infty_- <0$ increases monotonically as $\sigma_0$ increases.
The solid black line indicates the limit of the outer event horizon, Eq. \eqref{outerevent}, the dashed black line represents the limiting corotating circular photon orbit, Eq. \eqref{circularorbitphotonrad}, while the dash-dotted black line corresponds to the innermost stable circular orbit, Eq. \eqref{rmargbsc}. The limit $r/M = 2$ corresponds to the outer boundary of the ergosphere at $\theta = \pi/2$.}
\label{fig3}
\end{figure}

\begin{figure}[]
\begin{center}
\includegraphics[width=8.4cm]{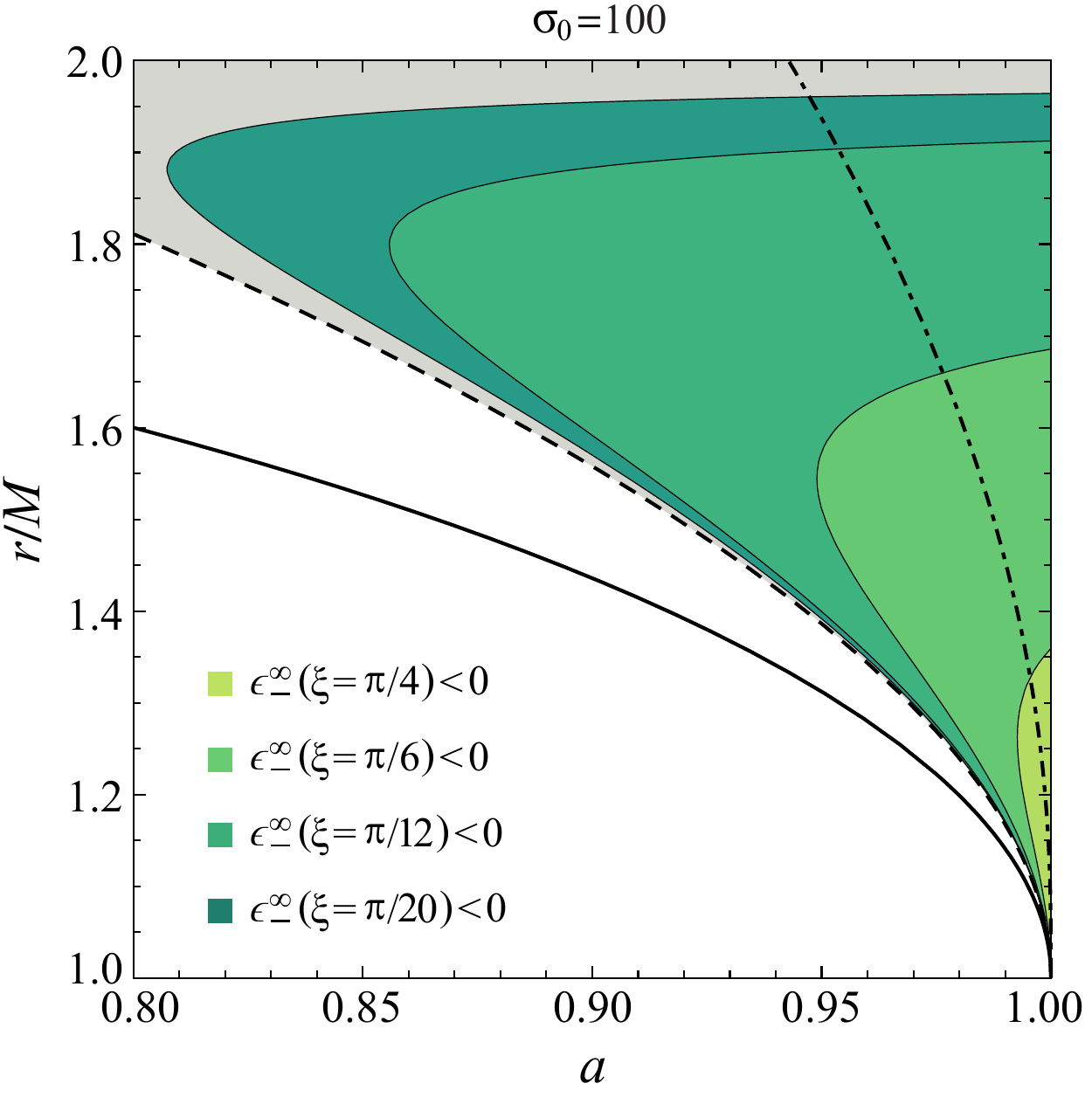}
\vspace{-0.30cm}
\end{center}
\caption{Regions of the  phase-space $\{a,r/M\}$ where the energies at infinity per enthalpy from Eq.  \eqref{energuisMagnet} are such that $\Delta \epsilon^\infty_+ >0$ (gray area) and $\epsilon^\infty_- <0$ (green areas), for plasma magnetization $\sigma_0 = 100$ and different values of the orientation angle $\xi \in \left\{ {\pi/20,\pi/12,\pi/6,\pi/4} \right\}$. Other lines are the same as in Figure \ref{fig3}. The area with $\epsilon^\infty_- <0$ increases monotonically as $\xi$ decreases.}
\label{fig4}
\end{figure}

We analyze the viability of energy extraction via magnetic reconnection by considering solutions of Eq. \eqref{energuisMagnet}. In particular, in Figs. \ref{fig3} and \ref{fig4} we display the regions of the phase-space $\{a,r/M\}$ where $\epsilon^\infty_- <0$ and $ \Delta \epsilon^\infty_+  >0$, which correspond to the conditions for energy extraction. This is done for a reconnecting magnetic field with orientation angle $\xi = \pi/12$ and different values of the magnetization parameter $\sigma_0 \in \left\{ {1,3,10,30,100} \right\}$ (Fig. \ref{fig3}), and for a plasma magnetization $\sigma_0 = 100$ and different values of the orientation angle $\xi \in \left\{ {\pi/20,\pi/12,\pi/6,\pi/4} \right\}$ (Fig. \ref{fig4}).

As the magnetization of the plasma increases, the region of the phase-space $\{a,r/M\}$ where magnetic reconnection extracts black hole rotational energy extends to larger $r/M$ values and lower values of the dimensionless spin $a$ (Fig. \ref{fig3}). From Eq. \eqref{energuisMagnet} we can see that $\epsilon^\infty_-$ is a monotonically decreasing function of $\sigma_0$, while $\epsilon^\infty_+$ monotonically increases with $\sigma_0$. $\epsilon^\infty_+ > 0$ is easily satisfied for $r_{\rm ph} < r < r_E$, $a>0$, and $\xi < \pi/2$. On the other hand, $\epsilon^\infty_- < 0$ requires $\sigma_0 \gg 1$ in order for reconnection to extract black hole energy in a significant region of the phase-space $\{a,r/M\}$. High values of the plasma magnetization can extend the energy extraction region up to the outer boundary of the ergosphere, while energy extraction for moderate values of the spin parameter $a$ is subject to the occurrence of particle orbits inside the ergosphere.

Energy extraction via magnetic reconnection is also favored by reconnection outflows whose orientation is close to the azimuthal direction. The region of the phase-space $\{a,r/M\}$ where energy extraction occurs increases to larger $r/M$ values and lower $a$ values as the orientation angle $\xi$ decreases. Notwithstanding, even an angle as large as $\xi = \pi/4$ admits a modest region of the phase-space where magnetic reconnection extracts rotational energy. The increase of the energy extraction region for decreasing angle $\xi$ is due to the fact that only the azimuthal component of the outflow velocity contributes to the extraction of rotational energy. For an angle $\xi = \pi/20$, the extraction of black hole energy happens for $X$-points up to $r/M \approx 1.96$ (for $\sigma_0 =100$), while $\xi \rightarrow 0$ can extend this margin up to the outer boundary of the ergosphere.

The ergosphere of spinning black holes ($r_{H} <r < r_{E}$) can reach very high plasma magnetizations (e.g, $\sigma_0 \gg 100$ close to the event horizon of the black hole M87* \cite{EHT_5_2019}). Furthermore, for rapidly spinning ($a$ close to unity) black holes, we expect a reconnecting magnetic field with small orientation angle, $\xi  \lesssim \pi/6$, as the strong frame-dragging effect inside the ergosphere stretches the magnetic field lines along the azimuthal direction \citep[e.g.][]{Koide02,Semenov04}. Therefore, the plots shown in Figs. \ref{fig3} and \ref{fig4} indicate that magnetic reconnection is a viable mechanism for extracting energy from rotating black holes with dimensionless spin $a$ close to unity. On the other hand, energy extraction via magnetic reconnection becomes negligible for spin values $a \lesssim 0.8$. The availability of reconnection regions inside the ergosphere decreases as the spin parameter decreases, with no circular orbits inside the ergosphere for spin $a \leq 1/\sqrt{2}$. Magnetic reconnection could still be capable of extracting energy in such cases if a circular orbit is sustained thanks to the help of the magnetic field or if one considers non-circular orbits.

\section{Energy Extraction Rate and Reconnection Efficiency} \label{section4}

We now evaluate the rate of black hole energy extraction. This depends on the amount of plasma with negative energy at infinity that is swallowed by the black hole in the unit time. Therefore, a high reconnection rate can potentially induce a high energy extraction rate. The power $P_{\rm extr}$ extracted from the black hole by the escaping plasma can be estimated as
\begin{equation} \label{Pextr}
P_{\rm extr} = - \epsilon_-^\infty w_0 A_{\rm in} U_{\rm in}  \, ,
\end{equation}
where $U_{\rm in} = \mathcal{O}(10^{-1})$ for the collisionless regime, while $U_{\rm in} = \mathcal{O}(10^{-2})$ for the collisional one. $A_{\rm in}$ is the cross-sectional area of the inflowing plasma, which can be estimated as ${A}_{\rm in} \sim (r_E^2 - r_{{\rm ph}}^2)$ for rapidly spinning black holes. In particular, for $a \rightarrow 1$ one has $(r_E^2 - r_{{\rm ph}}^2) = (r_{E}^2 - r_{H}^2) = 3M^2$.

\begin{figure}[]
\begin{center}
\includegraphics[width=8.4cm]{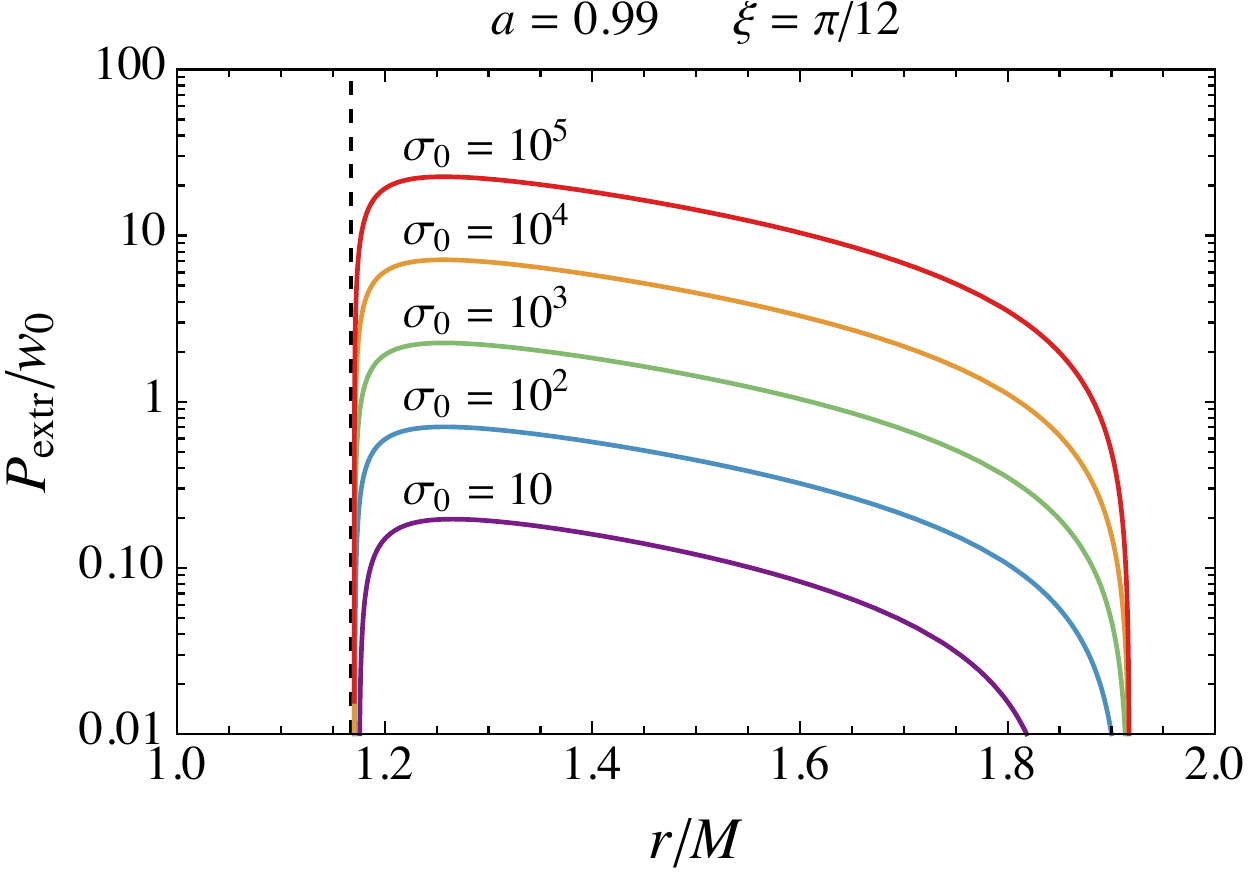} 
\bigskip $\,$
\hspace*{-0.05cm}\includegraphics[width=8.4cm]{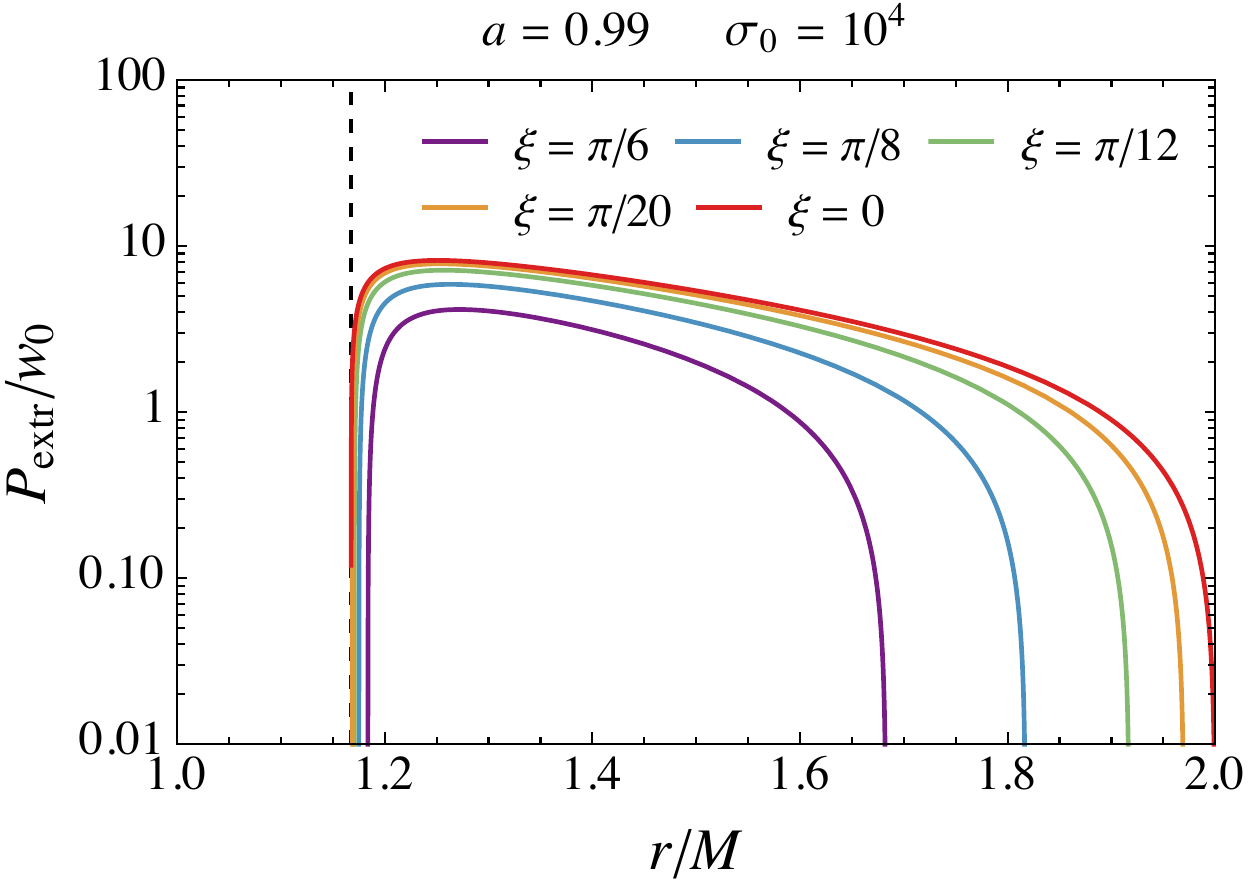}
\vspace{-0.30cm}
\end{center}
\caption{${P_{\rm extr}}/w_0 = - \epsilon_-^\infty A_{\rm in} U_{\rm in}$ as a function of the dominant $X$-point location $r/M$ for a rapidly spinning black hole with $a = 0.99$ and reconnection inflow four-velocity $U_{\rm in} = 0.1$ (i.e., collisionless reconnection regime). $\epsilon_-^\infty$ is evaluated using Eq. \eqref{energuisMagnet}, while $A_{\rm in} = (r_{{\rm ph}}^2 - r_{H}^2)$. We have also set $M=1$. Different colors (from indigo to red) refer to different plasma magnetizations (from $\sigma_0 = 10$ to $\sigma_0 = 10^5$) and $\xi = \pi/12$ (top panel) or different orientation angles (from $\xi = \pi/6$ to $\xi = 0$) and $\sigma_0 = 10^4$ (bottom panel). The vertical dashed line indicates the limiting circular orbit $r_{\rm ph}(a=0.99)$.}
\label{fig5}
\end{figure}

We show in Fig. \ref{fig5} the ratio $P_{\rm extr}/w_0$ as a function of the dominant $X$-point location $r/M$ for a rapidly spinning black hole with $a=0.99$ and magnetic reconnection in the collisionless regime. 
This is done for a typical reconnecting magnetic field with orientation angle $\xi = \pi/12$ and different values of the magnetization parameter $\sigma_0 \in \left\{ {10,10^2,10^3,10^4,10^5} \right\}$ (top panel), and for a typical magnetization $\sigma_0 = 10^4$ and different values of the orientation angle $\xi \in \left\{ {0,\pi/20,\pi/12,\pi/8,\pi/6} \right\}$ (bottom panel). 
The power extracted from the black hole increases monotonically for increasing values of the plasma magnetization and for lower values of the orientation angle. It reaches a peak for $X$-point locations that are close to the limiting circular orbit until it drops off. The peak of the extracted power can continue to raise up to a maximum value that is achieved for $r/M \rightarrow 1$ if $a \rightarrow 1$. The theoretical limit of the maximum power is given by 
\begin{equation} \label{PextrMAX}
P_{\rm extr}^{\rm max} \simeq  \sqrt{\sigma_0/3} \, w_0 A_{\rm in} U_{\rm in}  \sim 0.1 M^2 \sqrt{\sigma_0} \, w_0  \, ,
\end{equation}
which follows directly from Eqs. \eqref{energ_minus_simple} and \eqref{Pextr}. We can see from Fig. \ref{fig5} that the peak of the extracted power is already close to the maximum theoretical limit when $\xi \lesssim \pi/12$. 

\begin{figure}[]
\begin{center}
\includegraphics[width=8.4cm]{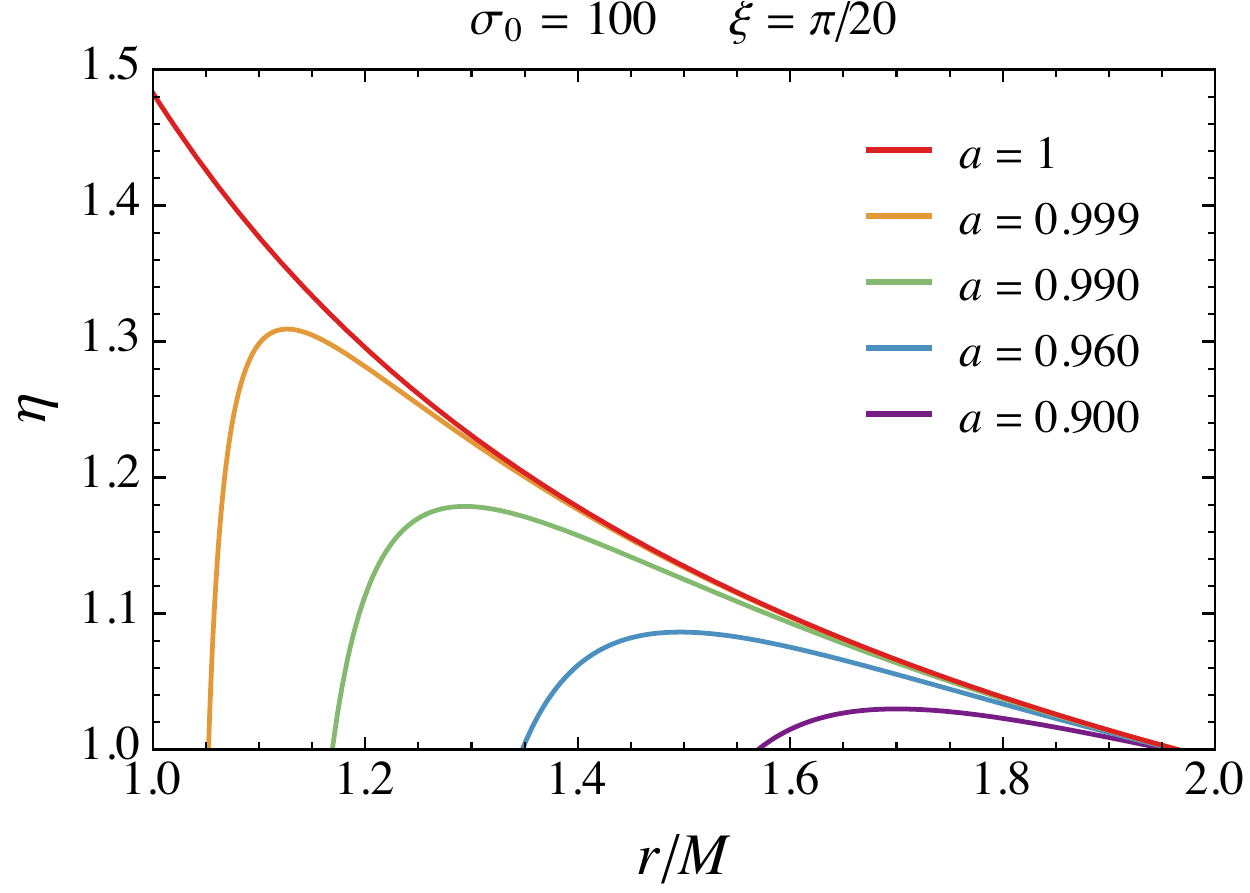}
\vspace{-0.30cm}
\end{center}
\caption{Efficiency $\eta$ of the reconnection process as a function of the dominant $X$-point location $r/M$ for a reconnection layer with upstream plasma magnetization $\sigma_0 = 100$ and reconnecting magnetic field having orientation angle $\xi = \pi/20$. Different colors (from indigo to red) refer to different black hole spin values (from $a = 0.9$ to $a = 1$). }
\label{fig6}
\end{figure}

The proposed mechanism of energy extraction via magnetic reconnection generates energetic plasma outflows that steal energy from the black hole, but it also necessitates magnetic field energy to operate. Magnetic energy is indeed needed in order to redistribute the angular momentum of the particles in such a way to generate particles with negative energy at infinity and particles escaping to infinity. Therefore, it is convenient to define the efficiency of the plasma energization process via magnetic reconnection as
\begin{equation}  \label{eff}
\eta =  \frac{\epsilon^\infty_+}{\epsilon^\infty_+ + \epsilon^\infty_-}   \,  .
\end{equation}
Extraction of energy from the black hole takes place when $\eta > 1$. Figure \ref{fig6} shows the efficiency $\eta$ as a function of the dominant $X$-point location $r/M$ for a reconnection layer with magnetization parameter $\sigma_0=100$, orientation angle $\xi = \pi/20$, and different black hole spin values $a \in \left\{ {0.90,0.96,0.99,0.999,1} \right\}$. The efficiency $\eta$ significantly increases for reconnection $X$-points that are closer to the black hole event horizon and falls off below unity when the inner radius reaches $r_{\rm ph}$. The maximum efficiency can be evaluated by considering the optimal energy extraction conditions ($a, r/M \rightarrow 1$, $\xi \rightarrow 0$) and $\sigma_0 \gg 1$. In this case, Eq. \eqref{eff} gives
\begin{equation}  \label{effmax}
\eta_{\rm max} \simeq \frac{\sqrt{3 \sigma_0}}{ \sqrt{3 \sigma_0} -  \sqrt{\sigma_0/3}} = {3}/{2}  \,  .
\end{equation}
Therefore, the additional energy extracted from the black hole, while non-negligible, do not extensively modify the energetics of the escaping plasma.

We can also compare the power extracted from the black hole by fast magnetic reconnection with the one that can be extracted via the Blandford-Znajek mechanism, in which the rotational energy is extracted electromagnetically through a magnetic field that threads the black hole event horizon.
For maximum efficiency conditions \cite{MT82,Thorne86,komissarov01}, the rate of black hole energy extraction via the Blandford-Znajek mechanism is given by \cite{BZ77,Tchekhovskoy10} 
\begin{equation} \label{P_BZ}
P_{\rm BZ} \simeq \kappa \Phi_{\rm BH}^2 \left( {\Omega_H^2 + \chi \Omega_H^4 + \zeta \Omega_H^6 } \right) \, ,
\end{equation}
where $\Phi_{\rm BH} = \frac{1}{2} \int_{\theta} \int_{\phi} |B^r| dA_{\theta \phi}$ is the magnetic flux threading one hemisphere of the black hole horizon (with $dA_{\theta \phi} = \sqrt{-g} \, d\theta d\phi$ indicating the area element in the $\theta$-$\phi$ plane), $\Omega_H = a /2 r_{H}$ is the angular frequency of the black hole horizon, while $\kappa$, $\chi$, and $\zeta$ are numerical constants. The numerical prefactor $\kappa$ depends on the magnetic field geometry near the black hole ($\kappa \approx 0.053$ for a split monopole geometry and $\kappa \approx 0.044$ for a parabolic geometry), while $\chi \approx 1.38$ and $\zeta \approx -9.2$ \cite{Tchekhovskoy10}. 
Eq. \eqref{P_BZ} is a generalization of the original Blandford-Znajek scaling \cite{BZ77} $P_{\rm BZ} \simeq \kappa \Phi_{\rm BH}^2 (a/4M)^2$, which is recovered in the small spin limit $a \ll 1$.

\begin{figure}[]
\begin{center}
\includegraphics[width=8.4cm]{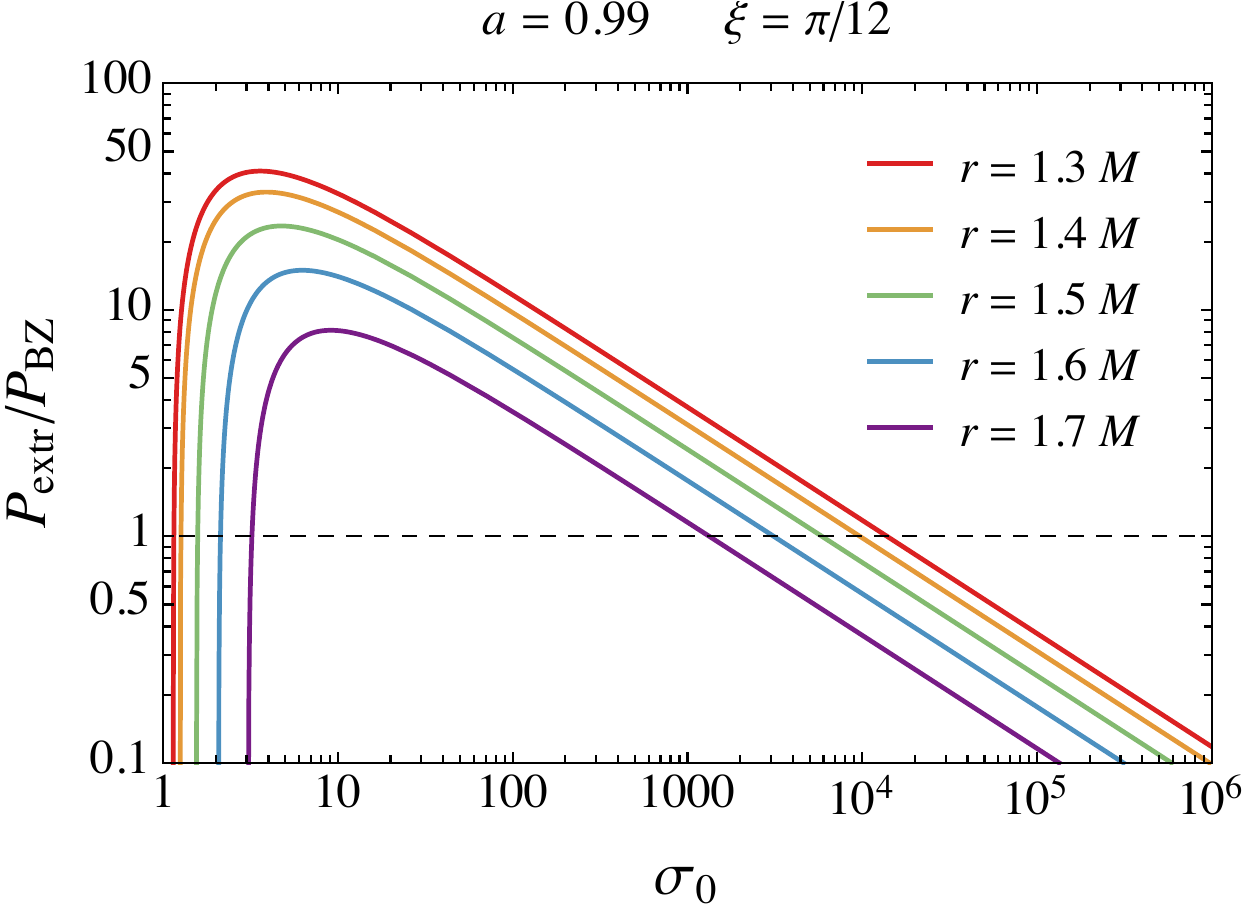}
\vspace{-0.30cm}
\end{center}
\caption{Power ratio ${P_{\rm extr}}/{P_{\rm BZ}}$ as a function of the plasma magnetization $\sigma_0$ for a black hole with dimensionless spin $a = 0.99$ and a reconnecting magnetic field having orientation angle $\xi = \pi/12$. Different colors (from indigo to red) refer to different dominant $X$-point locations $r/M \in \left\{ {1.3,1.4,1.5,1.6,1.7} \right\}$. We considered $U_{\rm in} = 0.1$ (i.e., collisionless reconnection regime), $A_{\rm in} = (r_{{\rm ph}}^2 - r_{H}^2)$, and $\kappa = 0.05$.}
\label{Fig7}
\end{figure}

In order to provide a rough order of magnitude estimate of the power extracted during the occurrence of fast magnetic reconnection with respect to the approximately steady-state Blandford-Znajek process,
we assume $\Phi_{\rm BH} \sim |B^r| r_{H}^2 \sim B_0 {\sin \xi} \, r_{H}^2$ (we point out that a precise evaluation of $\Phi_{\rm BH}$ requires direct numerical simulations that reproduce the detailed magnetic field configuration at all latitudes, while the angle $\xi$ is a good estimate for the magnetic field configuration only at low latitudes \citep[e.g.][]{Koide02,Semenov04}). Then, we can evaluate the ratio ${P_{\rm extr}}/{P_{\rm BZ}}$ as
\begin{equation} \label{powerratiowithBZ1}
\frac{P_{\rm extr}}{P_{\rm BZ}} \sim\frac{ - \epsilon_-^\infty A_{\rm in} U_{\rm in}} {\kappa \, \Omega_H^2 r_{H}^4 \sigma_0 \sin^2 \xi \, (1+ \chi \Omega_H^2 + \zeta \Omega_H^4)}\, .
\end{equation}
Fig. \ref{Fig7} shows the ratio ${P_{\rm extr}}/{P_{\rm BZ}}$ given by the right-hand side of Eq. \eqref{powerratiowithBZ1} as a function of the plasma magnetization $\sigma_0$ for the fast collisionless reconnection regime. ${P_{\rm extr}}/{P_{\rm BZ}} \gg 1$ for an extended range of plasma magnetizations. For $\sigma_0 \sim 1$, the force-free approximation (the inertia of the plasma is ignored, i.e. $w_0 \rightarrow 0$) that is used to derive the extracted power in the Blandford-Znajek process becomes invalid. In this case, magnetic reconnection is an effective mechanism of energy extraction provided that the plasma magnetization is sufficient to satisfy the condition $ \epsilon_-^\infty < 0$ (as well as $\Delta \epsilon^\infty_+ >0$). On the other hand, for $\sigma_0 \rightarrow \infty$, energy extraction via fast magnetic reconnection is always subdominant to the Blandford-Znajek process since ${P_{\rm extr}}/{P_{\rm BZ}} \rightarrow 0$ in this limit.
If we neglect higher order corrections with respect to  $\Omega_H^2$ (which leads to an overprediction of $P_{\rm BZ}$ by about 25\% as $a \rightarrow 1$ \cite{Tchekhovskoy10}), and recalling that $\Omega_H = 1/2M$ for $a \rightarrow 1$, we can estimate the ratio ${P_{\rm extr}}/{P_{\rm BZ}}$ for a rapidly spinning black hole as
\begin{equation}  \label{powerratiowithBZ2}
 \frac{P_{\rm extr}}{P_{\rm BZ}}\sim\frac{- \epsilon_-^\infty}{\kappa \,   \sigma_0 \sin^2 \xi}\, ,
\end{equation}
where we considered plasmoid-mediated reconnection in the collisionless regime. Therefore, the power extracted via fast collisionless magnetic reconnection can exceed the one extracted through the Blandford-Znajek process for an extended range of plasma magnetizations if there is a significant toroidal component of the magnetic field in the black hole ergosphere. Note, however, that energy extraction by fast magnetic reconnection is localized in time, since it requires a certain time to build-up the magnetic field configuration storing the magnetic energy that is eventually dissipated via fast magnetic reconnection.

\section{Conclusions}
\label{section5}

In this paper, we envisioned the possibility of extracting black hole rotational energy via fast magnetic reconnection in the black hole ergosphere. We considered a configuration with antiparallel magnetic field lines near the equatorial plane, which is induced by the frame dragging of the spinning black hole. The change in magnetic field direction at the equatorial plane produces an equatorial current sheet that is disrupted by the plasmoid instability when its aspect ratio reaches a critical value (for a collisionless relativistic pair plasma, the critical aspect ratio condition is derived in Ref. \cite{Comisso2019}). The formation of plasmoids/flux ropes drives fast magnetic reconnection, which rapidly converts the available magnetic energy into plasma particle energy. When the plasma is expelled out of the reconnection layer, the magnetic tension that drives the plasma outflow relaxes. The field lines are then stretched again as a consequence of the frame dragging and a current layer prone to fast plasmoid-mediated reconnection forms again. This process leads to reconnecting current sheets that form rapidly and intermittently.

Magnetic reconnection accelerates part of the plasma in the direction of the black hole rotation, while another part of the plasma is accelerated in the opposite direction and falls into the black hole. Black hole energy extraction occurs if the plasma that is swallowed by the black hole has negative energy as viewed from infinity, while the accelerated plasma that gains energy from the black hole escapes to infinity. Therefore, differently from the Blandford-Znajek process, in which the extraction of rotational energy is obtained through a purely electromagnetic mechanism, the energy extraction mechanism described here requires non-zero particle inertia. This mechanism is also different from the original Penrose process, since dissipation of magnetic energy is required to produce the negative-energy particles. Clearly, all mechanisms extract black hole rotational energy by feeding the black hole with negative energy and angular momentum.

We showed analytically that energy extraction via magnetic reconnection is possible when the black hole spin is high (dimensionless spin $a \sim 1$) and the plasma is strongly magnetized (plasma magnetization $\sigma_0 > 1/3$). 
Magnetic reconnection is assumed to occur in a circularly rotating plasma with a reconnecting field having both azimuthal and radial components. The region of the phase-space $\{a,r/M\}$ where magnetic reconnection is capable of extracting black hole energy depends on the plasma magnetization $\sigma_0$ and the orientation $\xi$ of the reconnecting magnetic field. We showed that high values of the plasma magnetization and mostly azimuthal reconnecting fields can expand the energy extraction region up to the outer boundary of the ergosphere. For a dimensionless spin parameter that approaches unity, the extraction of black hole energy is maximal when the dominant reconnection $X$-point (where the two magnetic reconnection separatrices intersect) is close to the event horizon. For $\sigma_0 \gg 1$, we showed that the asymptotic negative energy at infinity per enthalpy of the plasma that is swallowed by the black hole is $\epsilon^\infty_-  \simeq - \gamma_{\rm out} v_{\rm out}/ {\sqrt{3}} \simeq - \sqrt{\sigma_0/3}$. On the other hand, the plasma that escapes to infinity and takes away black hole energy asymptotes the energy at infinity per enthalpy $\epsilon^\infty_+  \simeq  \sqrt{3} \, \gamma_{\rm out} v_{\rm out}  \simeq \sqrt{3 \sigma_0}$.

We calculated the power extracted from the black hole by the escaping plasma and evaluated its maximum when the dominant reconnection $X$-point is close to the event horizon. This corresponds to $P_{\rm extr}^{\rm max} \sim 0.1 M^2 \sqrt{\sigma_0} \, w_0$ for the collisionless plasma regime and one order of magnitude lower for the collisional regime. The overall efficiency of the plasma energization process via magnetic reconnection can reach a maximum of $\eta_{\rm max}  \simeq 3/2$. Therefore, the additional energy extracted from the black hole, while important, do not extensively modify the energetics of the escaping plasma. On the other hand, the power extracted via fast magnetic reconnection can induce a significant reduction of the rotational energy of the black hole, ${d E_{\rm rot}}/{dt} = \epsilon_-^\infty w_0 A_{\rm in} U_{\rm in}$. This is effective when $a$ is close to unity. Therefore, if we consider a black hole with dimensionless spin parameter close to unity and define $\varpi = 1-a \ll 1$, we have ${d E_{\rm rot}}/{dt} \simeq - (M/4 \sqrt{\varpi}) d\varpi/dt$ and the spindown time can be obtained as
\begin{equation}  \label{}
{t_{\rm sd}} = \frac{\mathcal{O}(10)}{2 \sqrt{\sigma_0} \, w_0 M} (\sqrt{\varpi_{\rm f}}-\sqrt{\varpi_{\rm i}}) \, ,
\end{equation}
where the subscripts ${\rm f}$ and ${\rm i}$ are used to label final and initial values, respectively. This indicates that magnetic reconnection can cause a significant spindown of the black hole when $a \sim 1$. For example, fast magnetic reconnection in the ergosphere can reduce the black hole dimensionless spin from $a=0.999$ to $a=0.99$ in ${t_{\rm sd}} \sim 1/(\sqrt{\sigma_0} \, w_0 M)$. On the other hand, at lower spin values, especially for $a <0.9$, magnetic reconnection loses its efficacy as the plasma available in the ergosphere diminishes.

Various systems hosting a black hole are expected to have magnetization $\sigma_0 \gtrsim 1$ in the ergosphere.
For the typical conditions around supermassive black holes in active galactic nuclei (AGNs), the energy density of the electromagnetic field far exceeds the enthalpy density of the plasma and $\sigma_0 \sim 10^{4}$ or larger \cite{DoddsEden2010,Ponti17,EHT_5_2019} is foreseeable. Likewise, long and short gamma-ray bursts (GRBs) may have $\sigma_0 \sim 1$ or larger \cite{MacFadyen99,vanPutten99,Kiuchi15,Ruiz19} in the ergosphere (a central black hole is assumed). 
Under these magnetization conditions (in addition to $a \sim 1$), magnetic reconnection is capable of extracting energy from the black hole. For $\sigma_0 \sim 1 - 10^4$, we have shown that the bursty energy extraction rate occurring during fast magnetic reconnection can exceed the more steady energy extraction rate expected from the Blandford-Znajek mechanism. On the other hand, as the plasma magnetization increases, energy extraction via fast magnetic reconnection becomes always subdominant since it requires non-vanishing plasma inertia.

In the scenario proposed here, fast magnetic reconnection occurs rapidly and intermittently, so that the associated emission within a few gravitational radii from the black hole is expected to be bursty in nature. This bursty behavior of fast magnetic reconnection might be responsible for triggering flares in the vicinity of rotating black holes. Indeed, frequent X-ray and near-infrared flares are detected on a regular basis from the Galactic Center black hole Sgr A* \citep[e.g.][]{Baganoff01,Genzel03,Meyer08,Neilsen13}, and magnetic reconnection close to the black hole is often conjectured to induce these flares \citep[e.g.][]{DoddsEden2010,ripperda20,Dexter20}. Recent observations by the GRAVITY collaboration \cite{Gravity2018} have been able to pin down the motion of near-infrared flares originating near the last stable circular orbit of Sgr A*. 
Reconnection layers originate naturally in the ergosphere of rotating black holes and produce plasmoids/flux ropes that are filled with energized plasma with an energy budget that can exceed the energy originally stored in the magnetic field.

In this paper we have assumed that the plasma rotates circularly around the black hole. This assumption may be relaxed in order to treat more complex scenarios in which reconnection occurs in non-circular orbits. In this case, the plasma could approach the event horizon even when the black hole spin is not particularly high, expanding the parameter space region where magnetic reconnection can extract black hole energy. 
Another situation that could increase the efficacy of magnetic reconnection is the simultaneous presence of equatorial and non-equatorial current sheets \cite{ripperda20}, which may result in an increase of the extracted power to some degree.
Finally, for reconnecting magnetic fields that have a significant radial component, particle acceleration owing to the reconnection electric field can increase the rate of energy extraction and the overall efficiency of the reconnection process.

$\,$

\begin{acknowledgments}
We gratefully acknowledge discussions with Lorenzo Sironi, Daniel Gro\v{s}elj, Russell Kulsrud, Manasvi Lingam, Yi-Hsin Liu, Joonas N\"attil\"a, Kyle Parfrey, Bart Ripperda, Daniel Siegel, and Yajie Yuan. L.C. acknowledges support by the NASA ATP NNX17AG21G and NSF PHY-1903412 grants. F.A.A. acknowledges support by the Fondecyt-Chile Grant No. 1180139. 
\end{acknowledgments}


\begin{thebibliography}{}
\bibitem{note1} Here we do not consider Hawking's quantum mechanical spontaneous emission \cite{hawking74}, which is negligible for black holes of mass $\gtrsim 1 M_\odot$.
\end{thebibliography}
\end{document}